\documentclass[preprint,showpacs,preprintnumbers,amsmath,amssymb]{revtex4}

\usepackage{graphicx}
\usepackage{dcolumn}
\usepackage{bm}

\begin{document}

\preprint{MPIK / Ion storage}

\title{Chemical probing spectroscopy of H$_3^+$ \\ above the barrier to linearity}

\author{
Holger Kreckel\footnote{holger@astro.columbia.edu / Present address: Columbia University, 550 West 120th Street, New York, NY 10027, USA},
Dennis Bing,
Sascha Reinhardt,
Annemieke Petrignani,
Max Berg
and Andreas Wolf}
\affiliation{Max-Planck-Institut f\"ur Kernphysik, Saupfercheckweg 1, 69117 Heidelberg, Germany}

\date{\today}

\begin{abstract}
We have performed chemical probing spectroscopy of H$_3^+$ ions trapped in a cryogenic 22-pole ion trap. The ions were buffer-gas cooled to $\sim55$\,K by collisions with helium and argon. Excitation to states above the barrier to linearity was achieved by a Ti:Sa laser operated between 11\,300 and 13\,300\,cm$^{-1}$. Subsequent collisions of the excited H$_3^+$ ions with argon lead to the formation of ArH$^+$ ions that were detected by a quadrupole mass spectrometer with high sensitivity. We report the observation of 17 previously unobserved transitions to states above the barrier to linearity. Comparison to theoretical calculations suggests that the transition strengths of some of these lines are more than five orders of magnitude smaller than those of the fundamental band, which renders them -- to the best of our knowledge -- the weakest H$_3^+$ transitions observed to date.   
\end{abstract}

\maketitle

\section{Introduction}

\begin{figure}[b]
\center
\includegraphics[width=11.5cm]{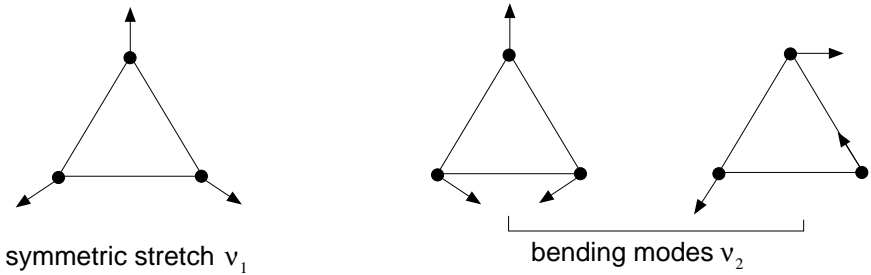}
\vspace*{-0.3cm}
\caption{Vibrational normal modes of H$_3^+$. The symmetric-stretch mode $\nu_1$ conserves the equilateral triangular symmetry and is infrared inactive, while the doubly-degenerate bending mode has a temporary dipole moment and is infrared active.}
\label{modes}
\end{figure}

Nearly one century after its discovery by J.J. Thomson in 1911 \cite{thomson11} the triatomic hydrogen ion is still challenging researchers all over the world. The outstanding importance of H$_3^+$ for the chemistry of interstellar matter and its presence in planetary atmospheres are well established \cite{tennyson03}, although the overabundance of H$_3^+$ in the diffuse interstellar medium is still not understood \cite{geballe06}. On the other hand, consisting of three protons and two electrons only, H$_3^+$ is the simplest polyatomic molecule and as such an important benchmark system for quantum mechanical calculations.
Spectroscopy of H$_3^+$ is complicated by the absence of a permanent dipole moment and by the fact that no stable electronically excited states are known. Consequently, the first laboratory spectra of H$_3^+$ that were observed by Oka in 1980 \cite{oka80} after a decades-spanning search, featured vibrational transitions in the infrared. H$_3^+$ is an equilateral triangle in its ground state and has two vibrational modes, the infrared-inactive symmetric stretch mode $\nu_1$ and the doubly-degenerate bending mode $\nu_2$ (see Fig.\,\ref{modes}). The latter posesses a small temporary dipole moment and hence is infrared active. All spectroscopic studies of H$_3^+$ performed to date rely on the $\nu_2$ mode, its overtones or $\nu_1$-$\nu_2$ combination bands; a comprehensive review of H$_3^+$ spectroscopy can be found in Ref.\,\cite{lindsay01}.

Owing to its fundamental relevance, triatomic hydrogen is a much-studied system -- experimentally as well as theoretically. Since the publication of the Meyer-Botschwina-Burton (MBB) potential energy surface \cite{meyer86} the low-energy regime of the H$_3^+$ potential is known with high precision. Today, the most sophisticated theoretical studies include adiabatic and relativistic corrections and reach an accuracy of 0.02\,cm$^{-1}$ near the equilibrium geometry \cite{jaquet98}. Calculations of energy levels based on the latest surfaces show excellent agreement with spectroscopically confirmed levels up to $\sim$9\,000\,cm$^{-1}$ above the ground state.

\begin{figure}[b]
\center
\includegraphics[width=13.5cm]{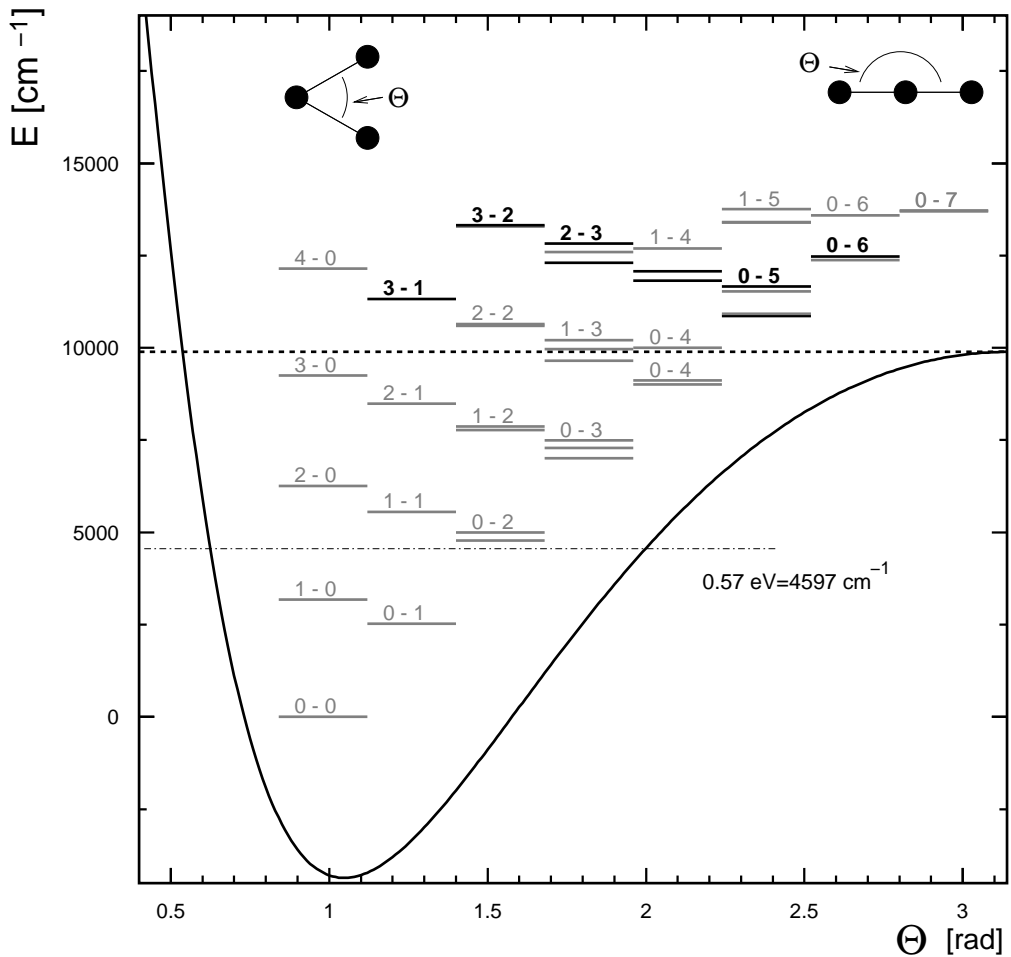}
\vspace*{-0.3cm}
\caption{The barrier to linearity. A special representation of the H$_3^+$ potential energy surface, where two H--H bonds are forced to be equal, while their length is varied to give the minimum energy for a given angle $\Theta$ between them. The plot shows that H$_3^+$ can sample linear geometries at energies $>$\,10\,000\,cm$^{-1}$ (dashed line). Vibrational levels $v_1$--$v_2$ with energies $<$\,14\,000\,cm$^{-1}$ are plotted and labeled. Those levels that have been observed in this work are emphasized in bold black color. The dash-dotted line marks the endothermicity of the ArH$^+$ formation reaction.  (Potential surface taken from Ref.\,\cite{prosmiti97}, zero point energy (level 0-0) set to 4\,362\,cm$^{-1}$ with respect to the potential minimum, according to Ref.\,\cite{roehse94}, level energies from Ref.\,\cite{alijah:private}). }
\label{pic:barrier}
\end{figure}

Nevertheless, for calculations of H$_3^+$ dissociation and dynamics, the energy range has to be extended to much higher energies. The construction of a so-called global potential energy surface \cite{prosmiti97}, however, turns out to be complicated since starting at 10\,000\,cm$^{-1}$ the H$_3^+$ molecule can sample linear geometries. This can readily be seen in Fig.\,\ref{pic:barrier}, showing a cut through the H$_3^+$ potential surface together with the vibrational levels. The high-energy region is challenging theory in particular since the basis sets optimized for triangular geometries near equilibrium are ill-suited to treat linear geometries \cite{tennyson00}. In an effort to create a surface that maintains high accuracy in the equilibrium region and includes the correct long-range behaviour at higher energies, Tennyson and coworkers employed a potential-switching approach in order to use the best-available calculations in each respective energy region \cite{prosmiti97,polyansky00}. 
Much of this theoretical research was motivated by the observation of a very rich H$_3^+$ photodissociation spectrum in the eighties \cite{carrington82,carrington93}. Carrington and coworkers detected more than 30\,000 discreet transitions starting out with rovibrationally excited H$_3^+$ and utilizing a CO$_2$ laser with a  limited wavelength range around $\sim$1\,000\,cm$^{-1}$. With the dissociation energy of H$_3^+$ being more than 35\,000\,cm$^{-1}$ \cite{cosby88} these experiments apparently probed an energy region very close to the dissociation limit. The origin of these lines as well as the underlying coarse structures are still unresolved \cite{tennyson00}.

One of the problems in this context is that no spectroscopic data exist to benchmark theory for almost two thirds of the path to dissociation (13\,000-35\,000\,cm$^{-1}$). Only recently have the first lines above the barrier to linearity been observed by the Oka group; so far 37 lines have been published, all lying in the wavelength region accessible with a Titanium:Sapphire laser (10\,700-13\,700\,cm$^{-1}$) \cite{gottfried06}. As the transition strengths of these lines are several thousand times weaker than the fundamental band, a very sophisticated highly optimized setup had to be employed to reach the required sensitivity \cite{gottfried03}. Furthermore, the presence of H$_2$ Rydberg states obscured some of the H$_3^+$ lines. In the latest publication by Gottfried \cite{gottfried06} the author concludes that ``the acquisition of additional transitions extending into the range of the visible dye laser ... will require a significant improvement in the experimental sensitivity''. In this work, we describe an experimental study that covers the same wavelength region accessible with a Ti:Sa laser system and we present 17 previously unpublished transitions. The experiment was performed using chemical probing spectroscopy in a cryogenic ion trap, a technique that has not yet reached its limits and demonstrates sensitivity reserves for potential application in the visible range of a dye laser and beyond.

This paper is organized as follows: in the next section a brief introduction into H$_3^+$ quantum numbers is given and theoretical contributions guiding us throughout the measurement are introduced. In Section\,3 the experimental method is explained; Section\,4 contains the results and in Section\,5 the perspectives of this work are presented.

\section{Theory overview}

\subsection{Quantum numbers}
The nomenclature used in this article follows the recommendations given in the review by Lindsay and McCall \cite{lindsay01}. Here, only a brief introduction in the labeling of levels and transitions and some basic selection rules will be given. For a more comprehensive representation the reader is referred to the original publication \cite{lindsay01} and references therein.

 Despite the fact that only parity and the total angular momentum ($F$) are good quantum numbers in a strict sense, we will introduce and use all vibrational and rotational quantum numbers -- although they are no longer unambiguous at higher energies  -- since these approximate quantum numbers are extremely helpful in understanding and classifying H$_3^+$ transitions.    

As seen before, the symmetric breathing vibration and the bending mode are labeled by $\nu_1$ and $\nu_2$, with integer quantum numbers $v_1$ and $v_2$, respectively. The bending mode is doubly degenerate and thus can host angular momentum. The corresponding quantum number is $\ell$ and its value ranges from $-v_2$ to $v_2$ in steps of two. The sign of $\ell$ is usually not specified and vibrational states are given by $v_1$$\nu_1+v_2\nu_2^{\,|\ell|}$. 

The motional angular momentum $J$ couples together with the nuclear spin angular momentum $I$ to the total angular momentum $F$. Two different values for $I$ exist, since the nuclear spin of the three protons that form H$_3^+$ can either add up to $I$\,=\,1/2 (para-H$_3^+$) or $I$\,=\,3/2 (ortho-H$_3^+$). The projection of the motional angular momentum $J$ onto the molecular symmetry axis is typically labeled $K$; for H$_3^+$, however, a new quantum number $g$ is introduced, which is defined by $g=K-\ell$. The energy of a given level is independent of the sign of $g$, and $G=|g|$ is used to define the energy levels, which have a strict double degeneracy for $G>0$.

\subsection{Selection rules}

Transitions between the two spin modifications of H$_3^+$ are highly forbidden \mbox{($\triangle I=0$)}. Therefore, for all spectroscopic purposes ortho- and para-H$_3^+$ can be treated as two separate entities and conservation of angular momentum then implies for dipole transitions
\begin{equation}
\triangle J = 0,\pm1\qquad \mbox{with} \qquad (J=0 \hspace{0.2cm}\not\hspace{-0.2cm}\longleftrightarrow J=0).
\end{equation}

Further selection rules arise from the symmetry of the molecule, a thorough derivation can be found in Ref.\,\cite{watson84}. Applying the permutation operations of the symmetry group of H$_3^+$ ($S_3^*$) onto its wave functions, one finds the following equations
\begin{eqnarray}
\triangle K&=&\,2\,i+1\,,\\
\triangle g&=& 3 \,i\,, 
\end{eqnarray}
with $i$ being an integer number ($i=0,\pm1,\ldots$).
The first term is a restatement of the requirement of parity change in dipole transitions as the parity of a H$_3^+$ wave function with a given $K$ is $(-1)^K$. The second selection rule is inherently connected to the conservation of nuclear spin since the following relations hold \cite{watson84}
\begin{eqnarray}
\mbox{ortho-H}_3^+ & \Longleftrightarrow & G=0~ mod~3 \label{ortho-g}\,,\\ 
\mbox{para-H}_3^+ & \Longleftrightarrow & G\ne0 ~ mod ~ 3 \label{para-g}\,.
\end{eqnarray} 
Owing to the absence of a permanent dipole moment, pure rotational transitions are strongly supressed. However, theoretical calculations \cite{pan86, miller88} predict that distortions due to the molecules rotation could lead to a {\it forbidden} rotational spectrum that has yet to be observed. 
Vibrational transitions between two breathing mode states are forbidden for the dipole moment does not change. Bending mode transitions on the other hand are allowed; in a harmonic picture, however, only excitations of a single quantum ($\triangle v_2=1$) are admitted  \cite{tennyson95} whereupon also the corresponding angular momentum quantum number $\ell$ has to change by unity ($\triangle \ell=1$). In reality, the H$_3^+$ ion is a rather anharmonic molecule and the latter selection rule is often violated, as will be demonstrated by the experimental results presented in Section 4.

\subsection{Transition labels}

All the transitions considered in this work start from the ground vibrational state, therefore it is sufficient to label the vibrational band of the upper state
\begin{equation}
v_1\nu_1+v_2\nu_2^{\,|\ell|}\,. 
\end{equation}
We label our observed transitions following the conventions established
in Refs.\,\cite{lindsay01,gottfried03} by
\begin{equation}
^{[n|t|\pm 6|\pm 9|... ]} \{ P|Q|R\}(J,G)^{[u|l]}_{[u|l]}\,.
\end{equation}
As the directions described in the literature need to be very
carefully interpreted for some transitions to highly mixed levels
occurring here, we choose to give a complete statement of the practical
rules applied in our transition labeling once the lower and the upper
level have been identified using the resources given in Sec II D. First, $\{P|Q|R\}$ refer to Delta $J=
{-1,0,+1}$ and ($J,G$) identify the corresponding quantum numbers of the
lower level.  Next, the superscripted prefix identifies the change in $G$.
No such prefix is applied when the two energy levels have the same $G$.  A
superscript $t$ is used whenever $\triangle$$G=+3$, while a superscript $n$ is
used when $\triangle$$G=-3$ and the $G$ value of the lower level is $\ge$$3$, or
when $|\triangle G|=1$ and the lower level has $G$$<3$; thus, these superscripts
represent transitions for which the (minimum allowed) change in $g$ is $|\triangle g|=3$
and where $\triangle g=\pm 3\,\mbox{sign}(g)$ for $t$ and $n$, respectively.  (Here,
sign($g$) refers to $g$ for the lower level.)
Similarly, a superscript $+6$
is used whenever $\triangle G=+6$, while a superscript $-6$ is used when $\triangle G=-6$
 and the $G$ value of the lower level is $\ge 6$, or when $|\triangle G|=
2,4$ and the lower level has $G<6$; thus, these superscripts represent
transitions for which the (minimum) change in $g$ is $|\triangle  g|=6$ and where
$\triangle  g=\pm 6\,\mbox{sign}(g)$ for the superscripts $+6$ and $-6$, respectively.
Higher transition orders with $|\triangle  g|=9,12,15\ldots$ may be indicated
correspondingly with indices $\pm|\triangle  g|$.  Note that some transitions
with a given $\triangle  G$ can in principle be realized by transitions of
various orders $|\triangle  g|$ (such as $|\triangle  g|=0$ and $|\triangle  g|=6$ for $G=3$
and $\triangle  G=0$); hence, for the purpose of the prefix labeling, we here
assume the minimum $|\triangle  g|$ required for a given combination of $G$'s.
Finally indices behind the ($J,G$) labels have to be applied in some
cases, since different combinations of $K$ and $\ell$ can lead to two
energetically different levels  with the same $G=|K-\ell |$ \cite{lindsay01}.  The
indices $[u|l]$ then distinguish the energetically upper ($u$) or lower ($l$)
one of a pair of levels with the same $G$ and are applied as a superscript
if required for the upper state and as a subscript if required for the
lower state.

\subsection{Calculations of H$_3^+$ levels and transitions} 
  
Many calculations of H$_3^+$ energy levels have been published, and it is beyond the scope of this article to review them all (for critical evaluations see Refs.\,\cite{lindsay01, gottfried03, gottfried06}). Of outstanding importance for the choice of observable lines was the comprehensive H$_3^+$ line list of Neale, Miller and Tennyson (NMT) \cite{neale96}. This list comprises all transitions between H$_3^+$ states up to 15\,000\,cm$^{-1}$. Besides the transition energy and the angular momentum and energy of the upper and lower levels, the Einstein $A$ coefficient for spontaneous decay is given for each transition, which allowed us to select the lines that might be strong enough to be observed experimentally.  

Another extremely valuable resource were the calculations of Schiffels, Alijah and Hinze (SAH) \cite{schiffels03b}. The critical evaluation of Gottfried \cite{gottfried06} showed that those SAH values that were spectroscopically adjusted have the smallest average error when compared to observed lines. As the NMT linelist lacks the individual rovibrational quantum numbers, the full labeling given for all levels by SAH was indespensable for the assignment of the transitions. Furthermore, additional assignments and corrections for levels based on the SAH code were provided by A. Alijah prior to publication \cite{alijah:private}.

\section{Experimental setup and procedure}

The experiment has been performed in a cryogenic ion trap ensuring buffer gas cooling of the H$_3^+$ ions to the lowest rotational levels. 
Due to the weakness of H$_3^+$ transitions and the comparatively low ion densities it is not practical to do conventional absorption spectroscopy inside the cryogenic ion trap. As a much more sensitive approach laser-induced chemical reactions can be used to probe photon absorption, since the reaction products can be detected with near-unity efficiency. First chemical probing experiments in an ion trap were conducted by Schlemmer {\it et al.} for the charge exchange reaction between N$_2^+$ and neutral Ar \cite{schlemmer99}. In the present work we employed the endothermic reaction
\begin{equation}
{\rm H}_3^+ + {\rm Ar} \quad\longleftrightarrow\quad {\rm ArH}^+ + {\rm H} \quad (-\,0.57\,{\rm eV})\,,
\label {eq:arhform}
\end{equation}
which has been used in a previous study to measure the population of single rotational states inside the cryogenic ion trap by inducing vibrational transitions of the third overtone of the bending mode (around $\sim$7\,200\,cm$^{-1}$) with a diode laser \cite{mikosch04, petrignani08}. In Fig.\,\ref{pic:barrier} the energy required for this reaction is marked in the level scheme by a dash-dotted line at $\sim$4\,600\,cm$^{-1}$. As can be seen, any H$_3^+$ state with more than one vibrational quantum ([$v_1$+$v_2]>1$) has enough energy to enable the formation of ArH$^+$.

\subsection{Setup}

\begin{figure}[t]
\center
\includegraphics[width=16cm]{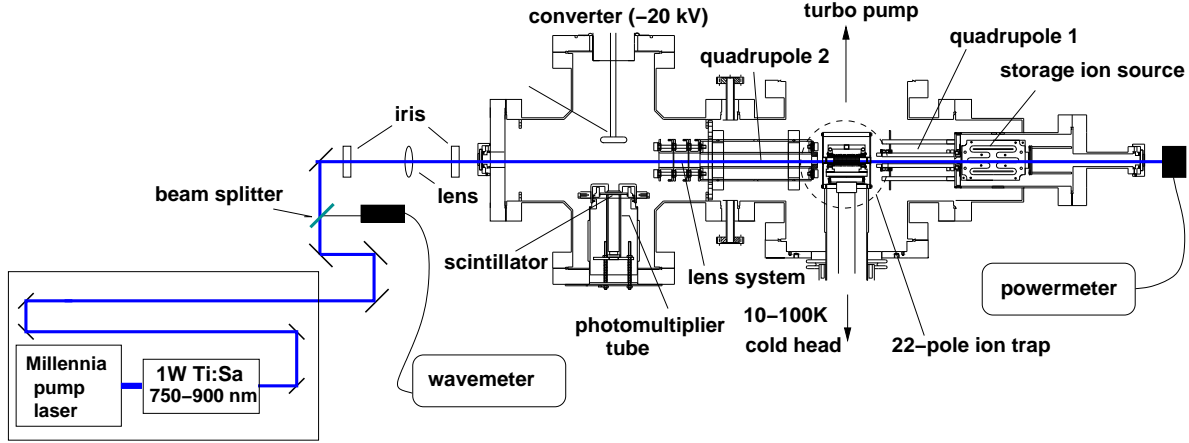}
\vspace*{-0.3cm}
\caption{Sketch of the trap setup and the laser system.}
\label{pic:setup}
\end{figure}

The heart of the setup is a 22-pole radiofrequency (RF) ion trap \cite{gerlich95, mikosch07} that is mounted on the second stage of a 10\,K cryogenic cold head (see Fig.\,\ref{pic:setup} for an overview). The setup was developed and used as an ion injector \cite{kreckel05b} for the TSR storage ring  to facilitate dissociative recombination measurements with rovibrationally cold ions \cite{kreckel05}. 

 The 22-pole trap itself is a cylindrical cage of 22 stainless steel rods of 1\,mm diameter and 40\,mm length that are alternatingly planted into two copper side plates. By applying an appropriate RF voltage (19.2\,MHz, $\sim$30\,V) to the rods, ions can be confined radially, while two small DC end electrodes placed inside bores in the side plates are used for axial ion confinement and for loading and unloading the trap by appropriate pulsing. The principle of the 22-pole trap is that of the Paul trap with the significant difference that a large field-free region in the trap center reduces radiofrequency heating to a minimum and hence -- in combination with a suitable buffer gas -- provides for distinctly lower ion temperatures. To this end, precooled He and Ar gas is let into the trap through openings in the copper ground plate.  The trap is encased in a massive copper housing to contain and cool the buffer gas while an aluminium heat shield mounted to the first stage of the cold head at $\sim$\,40\,K protects inner parts from the surrounding 300\,K blackbody radiation. For the chemical probing scheme Ar is needed as a reactant, therefore the trap had to be heated to 55\,K during the measurements by a heating pad underneath the ground plate, in order to prevent the Ar gas from freezing to the trap walls. The trap temperature can be determined by two calibrated diodes fixed to the copper housing. The vacuum chamber is pumped down to a base pressure on the order of 10$^{-8}$\,mbar by two turbomolecular pumps. At a trap temperature of 55\,K typical residual gas limited H$_3^+$ lifetimes of longer than 30\,s are achieved.

The H$_3^+$ ions are produced in a storage ion source \cite{teloy74} from normal hydrogen (n-H$_2$) gas and transported into the ion trap through a small RF quadrupole ion guide. Following storage inside the 22-pole trap the ions are extracted and mass selected by a second quadrupole in combination with a Daly type scintillation detector with near-unity efficiency \cite{daly60}. The axis that aligns the ion guides, the 22-pole trap and the storage ion source is kept free and vacuum viewports on each end are used to pass the laser beam through the entire setup (see Fig.\,\ref{pic:setup}).

The laser system consists of a continuous wave Coherent 899 Titanium:Sapphire laser that is pumped by a 10\,W Spectra-Physics Millennia frequency-doubled Nd:YVO$_4$ laser at 532\,nm. Two different sets of mirrors in the Ti:Sa ring oscillator were needed in order to cover the frequency range from 11\,200\,cm$^{-1}$ to 13\,300\,$^{-1}$ and the typical output power was between 0.4-1\,W. The frequency was determined by a commercial wavemeter that was calibrated by Doppler-free Rb spectroscopy with two hyperfine structure lines at $\sim$12\,816\,cm$^{-1}$. We assume an absolute uncertainty of 0.01\,cm$^{-1}$ for all transition frequencies.

\subsection{Measurement procedure}

Typically, a sample of 10$^3$ H$_3^+$ ions is produced in the storage ion source and transported into the 22-pole trap which is kept at (55\,$\pm$\,2)\,K throughout the measurement. Argon and helium gases are bled into the trap continuously at number densities of $n_{Ar}\sim10^{12}$\,cm$^{-3}$ and $n_{He}\sim 2\times 10^{14}$\,cm$^{-3}$. Whereas the He number density is kept high enough to allow for effective cooling, Ar is used as a probe gas to form ArH$^+$ ions. Hence, it is important to establish a high enough Ar-H$_3^+$ collision rate to make sure that an excited H$_3^+$ ion finds an Ar atom before the excited state decays spontaneously or is cooled in collisions with He. Assuming a Langevin collision rate coefficient of 1$\times 10^{-9}$\,cm$^{-3}$\,s$^{-1}$, the average time between collisions with Ar is on the order of 1\,ms which is significantly shorter than the average time it takes any excited state with at least three vibrational quanta to decay below the ($v_1$+$v_2)>1$ threshold \cite{mikosch04} and much shorter than the time needed to cool vibrational excitations in collisions with the He buffer gas.

 The hydrogen pressure in the ion source is kept as low as possible, however, hydrogen gas leaking out of the source aperture leads to a partial H$_2$ pressure of $2\times 10^{-7}$\,mbar in the trap vacuum chamber.
Furthermore, due to the low pressure in the ion source, many of the extracted H$_3^+$ ions are highly excited and react with Ar inside the trap immediately. The fraction of ions that enter the 22-pole with enough internal energy to form ArH$^+$ is found to be up to 60\,\%. The lifetime of ArH$^+$ ions inside the trap, however, is limited by the presence of H$_2$ initiating the backward reaction of Eq. (\ref{eq:arhform}). Typical ArH$^+$ lifetimes of $\tau_{ArH^+}=30$\,ms together with a rate coefficient of $2\times 10^{-9}$\,cm$^3$\,s$^{-1}$  \cite{bedford90} for the inverse of reaction (\ref{eq:arhform}) suggest a H$_2$ number density of $\sim$1.6$\times 10^{10}$\,cm$^{-3}$.

\begin{figure}[t]
\center
\includegraphics[width=11.7cm]{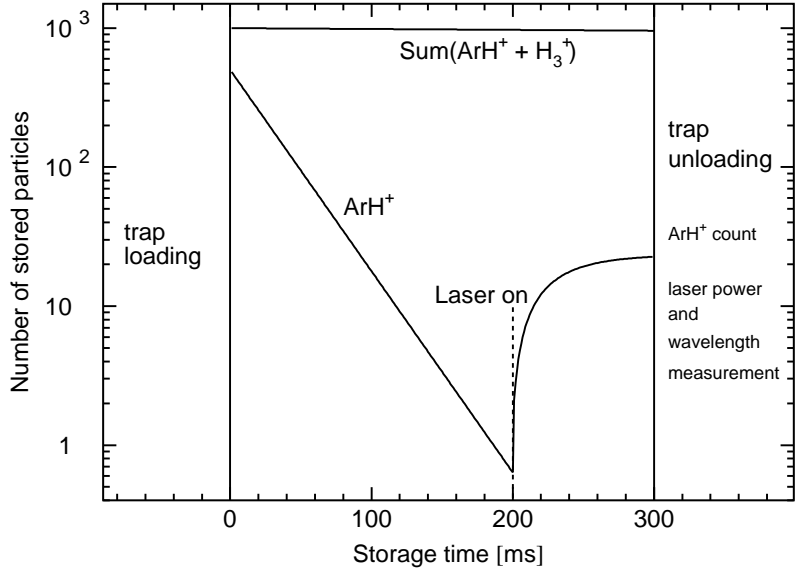}
\vspace*{-0.3cm}
\caption{Timing scheme and typical particle numbers of the measurements. Upon loading of the trap up to 60\,\% of the injected H$_3^+$ ions are highly exited and react to form ArH$^+$ immediately. After 200\,ms of storage the average number of ArH$^+$ ions has decayed below 1 due to the inverse of the reaction shown in Eq.\,(\ref{eq:arhform}). At 200\,ms the laser is activated for 100\,ms and laser-induced ArH$^+$ ions are created. All ions are extracted after 300\,ms and the ArH$^+$ ions are selected by a mass filter and counted. After each cycle consisiting of 20-100 trap fillings at fixed frequency the laser power and wavelength are determined. For the next cycle the laser wavelength is increased until a full wavelength scan is completed. Typically 5-20 wavelength scans were taken for each transition and the H$_3^+$ number (after 300\,ms of storage without laser interaction) and lifetime as well as the ArH$^+$ lifetime are recorded after each complete scan.}
\label{pic:scheme}
\end{figure}

After 200\,ms of storage, the average number of ArH$^+$ ions inside the trap has dropped by more than two orders of magnitude due to collisions with H$_2$ (see Fig. \ref{pic:scheme}). Now the laser beam is switched on for 100\,ms in order to saturate the yield of ArH$^+$ ions against their 30\,ms lifetime. The rate of ArH$^+$ production is given by
\begin{equation}
R_{ArH^+}(\nu)= \int_{\rm ovl} n_{H_3^+}\,f_{J,G}(\nu)\,B_{12}\,\rho(\nu)\,dV\, ,
\end{equation}
where $n_{H_3^+}$ is the H$_3^+$ number density inside the trap, $f_{J,G}(\nu)$ is the fraction of H$_3^+$ ions in a rovibrational state that can be excited by a given laser frequency $\nu$, $B_{12}$ is the Einstein coefficient for that transition and $\rho(\nu)$ is the spectral energy density. The integral runs over the interaction volume which is defined by the spatial overlap between the laser beam and the ion cloud inside the trap. The laser bandwidth of $\triangle \nu_L\sim$1\,MHz is large compared to the homogeneous line width, but small compared to the temperature-induced Doppler width inside the trap ($\sim$500\,MHz); the spectral energy density can be calculated 
\begin{equation}
\rho(\nu)=\frac{P_L}{\pi\,r_L^2\,c\,\triangle \nu_L}\quad,
\end{equation} 
with $P_L$ being the laser power and $r_L$ the radius of the laser beam. The Einstein coefficient can be derived from the calculated Einstein $A_{21}$ coefficients \cite{neale96} for spontaneous decay
\begin{equation}
B_{12}=\frac{2J_2+1}{2J_1+1}\,\,\frac{c^3}{8\pi h \nu^3}\,A_{21}\,.
\end{equation}
The fraction $f_{J,G}(\nu)\triangle\nu_L$ of H$_3^+$ ions that is addressed by the laser at a certain frequency $\nu$ is given by the population fraction of the ions in the lower state of the respective transition $f_{jg}$ multiplied with the value of the appropriate Doppler distribution for that frequency $f_{D}(\nu)$. At a trap temperature of 55\,K  more than 80\,\% of the H$_3^+$ ions are expected to be in one of the two lowest rovibrational states which differ in their nuclear spin configuration, with \mbox{($J$=1, $G$=1 /para)} and \mbox{($J$=1, $G$=0 /ortho)}, respectively. Although the para-state ($J$=1, $G$=1) is $\sim$33\,K lower in energy we assume both states to be populated equally since the high temperature equilibrium of the spin states realized in the n-H$_2$ ion source gas suggests a 1:1 ratio and we do not expect the nuclear spin temperature to change significantly in collisions with He or Ar (or the influx of H$_2$ from the source) inside the trap. Hence, we estimate each of these states to account for 40\,\% of the total population  ($f_{10}=f_{11}=0.4$). Only transitions starting from either one of these two states have been used for the chemical probing experiment.

The Doppler distribution can be described by
\begin{equation}
f_{D}=\frac{1}{\sqrt{2\pi}\sigma_D} e^{-(\nu-\nu_0)^2/2\sigma_D}\,,
\end{equation}
with the Doppler width given by
\begin{equation}
\sigma_D=\sqrt{\frac{k_B\,T}{m\,c^2}}\,\nu_0\,.
\end{equation}

To calculate the H$_3^+$ density we assume that 10$^3$ ions are distributed homogeneously over a cylindrical trap volume of $\pi \times (0.4$\,cm)$^2 \times 3$\,cm -- neglecting space charge and the influence of the endcaps \cite{trippel06} for the sake of simplicity -- which results in n$_{H_3^+}$$=660$\,cm$^{-3}$. The laser beam had a typical power of 0.5\,W with a radius of $r_L$=0.15\,cm resulting in a power density of $\rho=2.4\times 10^{-10}$\,W/cm$^3$. 
For an exemplary transition $5\nu_2^1\leftarrow 0$, $R(1,0)$ at $\nu_0$=11\,228.598\,cm$^{-1}$ with an Einstein $A_{21}$ coefficient of 1.018\,s$^{-1}$ ($B_{12}$=$7.2\times 10^{19}$\,cm$^3$/Js$^2$) and a Doppler width of $\sigma_D=514$\,MHz, the maximum ArH$^+$ formation rate in the center of the distribution is 
\begin{equation}
R_{ArH^+}=n_{H_3^+}\times f_{10} \times f_D(\nu_0)\times B_{12}\times \rho(\nu_0)\times\triangle\nu_L \times (\pi \,r_L^2\,\,d_{\rm eff})\,\,= 740\,{\rm s}^{-1},
\label{eq:formation}
\end{equation}
with a cylindrical overlap volume defined by the laser radius $r_L$ and the effective trap length $d_{\rm eff}=$3\,cm. 

The number of ArH$^+$ ions created per laser interaction time $t_L$ is
\begin{equation}
\frac{dN_{ArH^+}}{dt_L}=R_{ArH^+}-\frac{1}{\tau_{ArH^+}}\,N_{ArH^+},
\end{equation}
which is solved by
\begin{equation}
N_{ArH^+}=R_{ArH^+}\,\tau_{ArH^+}\,(1-e^{-t_L/\tau_{ArH^+}}),
\label{eq:diffsolve}
\end{equation}
and for long interaction times ($t_L \rightarrow \infty$) it leads to a steady state number of $\sim$22 ArH$^+$ ions per trap filling. Within the approximations made this is consistent with the value of 27.5 ions measured on this transition. For most of the weaker lines, however, we have increased the number of stored H$_3^+$ ions to several thousand to achieve a higher signal rate. In the present detection scheme this goes at the expense of reliably counting the number of stored ions which is needed for normalization. As the extracted H$_3^+$ particles come out of the trap in bunches of 30\,$\mu$s width, the detector (running in particle-counting mode) begins to show signs of saturation for more than a few hundred ions per pulse. Note that this problem arises {\it only} for the determination of the number of stored H$_3^+$ ions while the ArH$^+$ signal -- yielding a much lower count rate -- is unaffected. Hence, the saturation presently hampers the reliable measurement of relative line intensities. In future experiments, it is planned to use a faster photomultiplier and account for non-linearities in the normalization scheme in order to assess also the line intensities.

\begin{figure}[t]
\center
\begin{minipage}[t]{7cm}
\includegraphics[width=7cm]{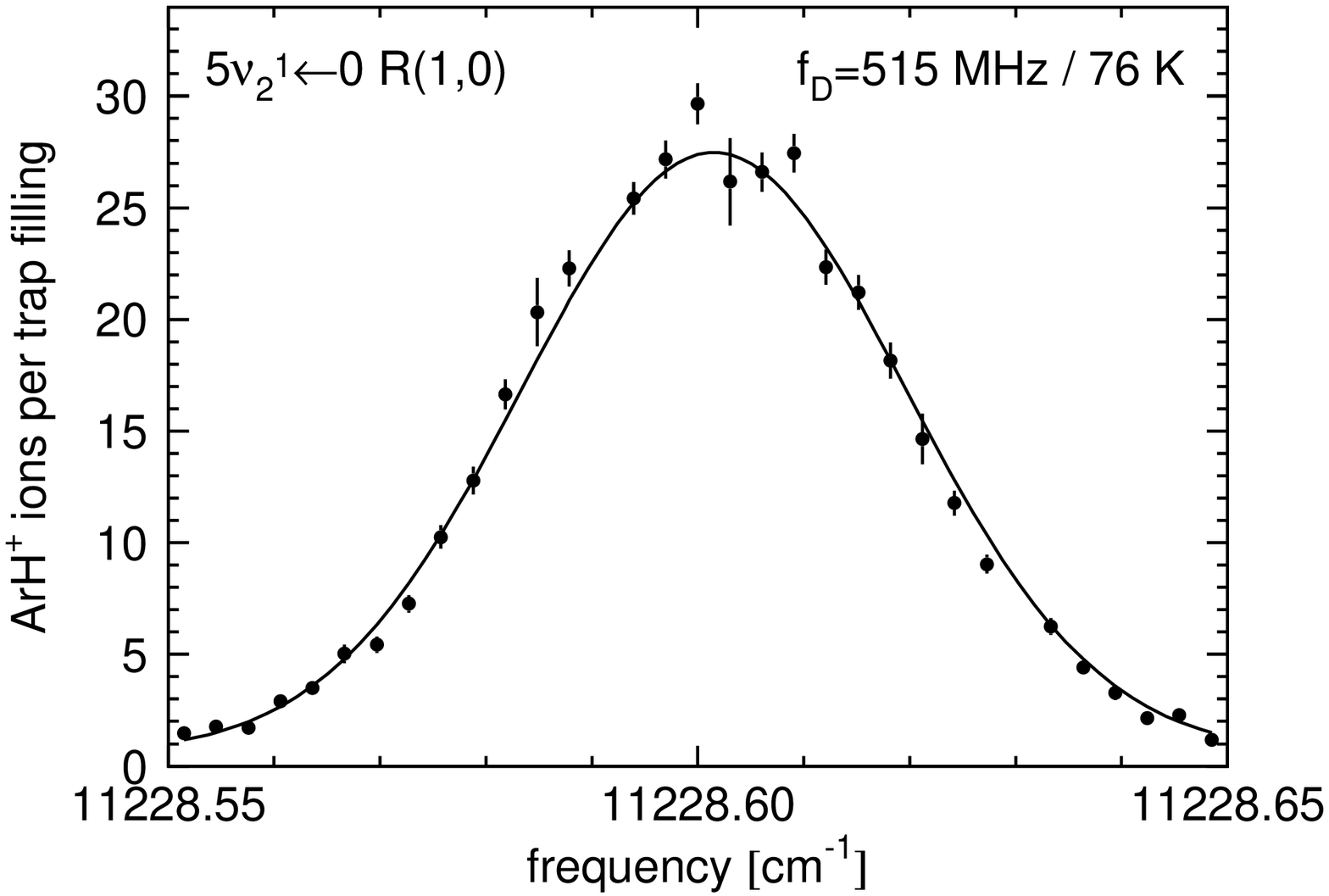}
\end{minipage}
\begin{minipage}[t]{7cm}
\includegraphics[width=7cm]{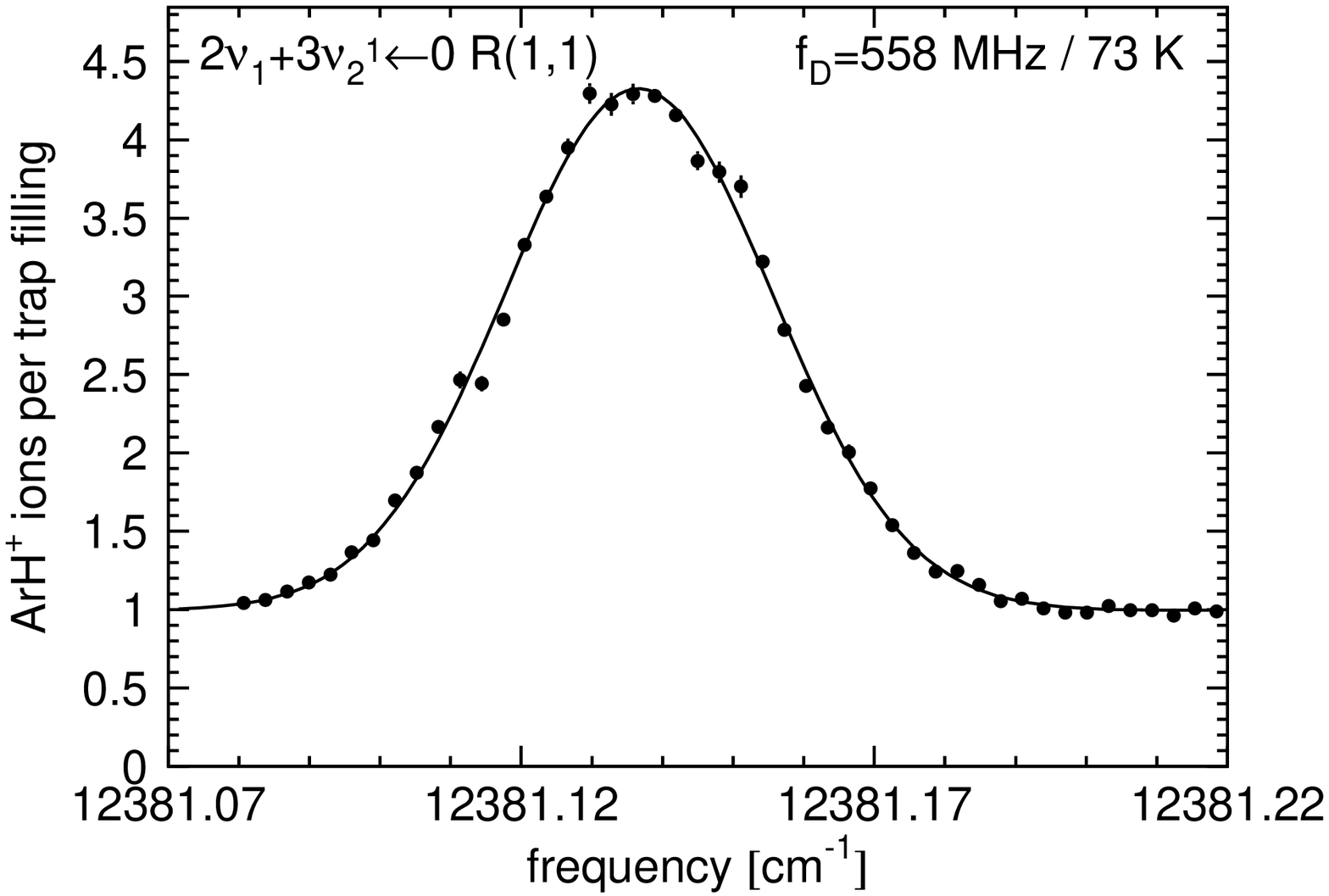}
\end{minipage}
\vspace*{-0.5cm}

\begin{minipage}[t]{7cm}
\includegraphics[width=7cm]{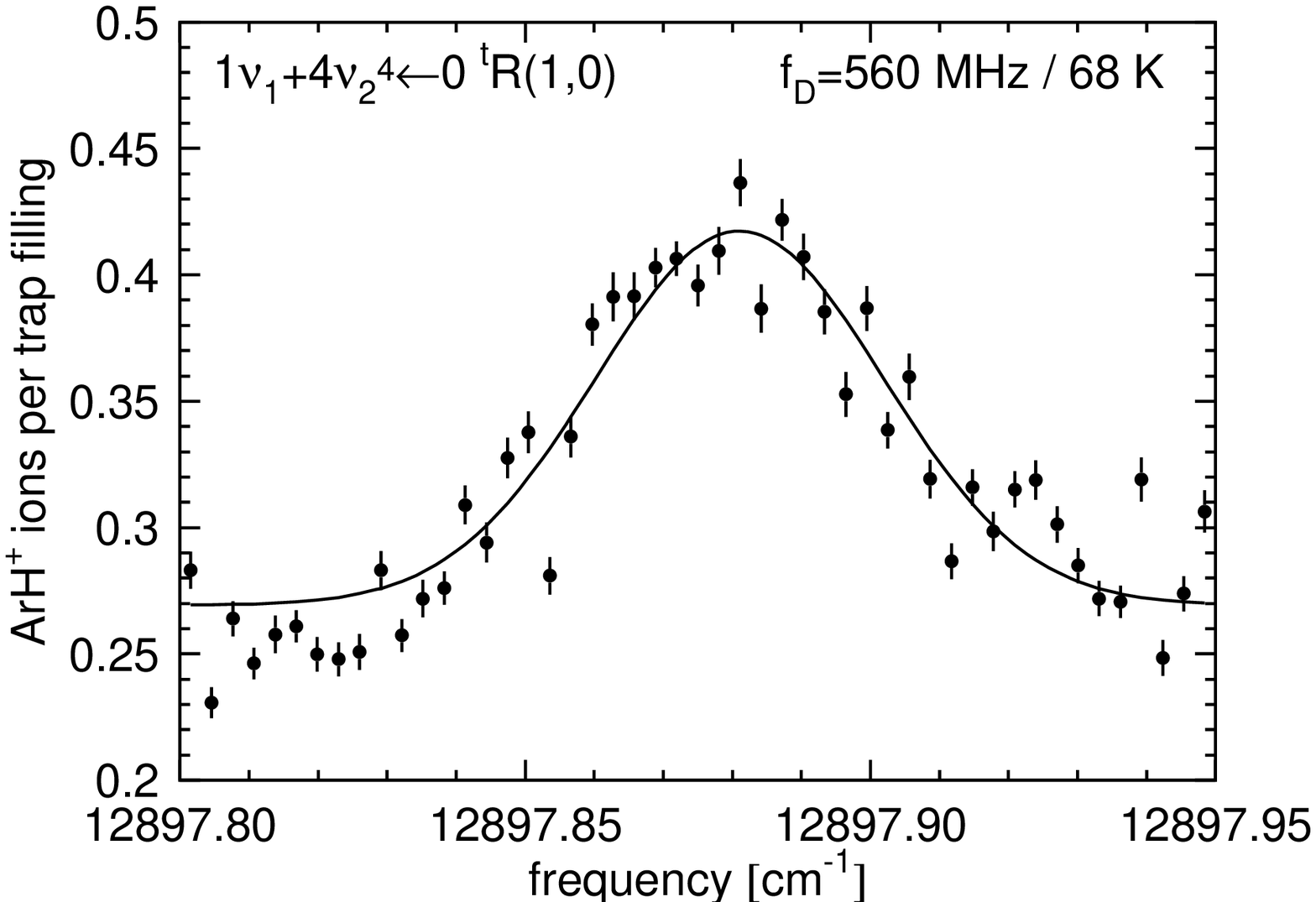}
\end{minipage}
\begin{minipage}[t]{7cm}
\includegraphics[width=7cm]{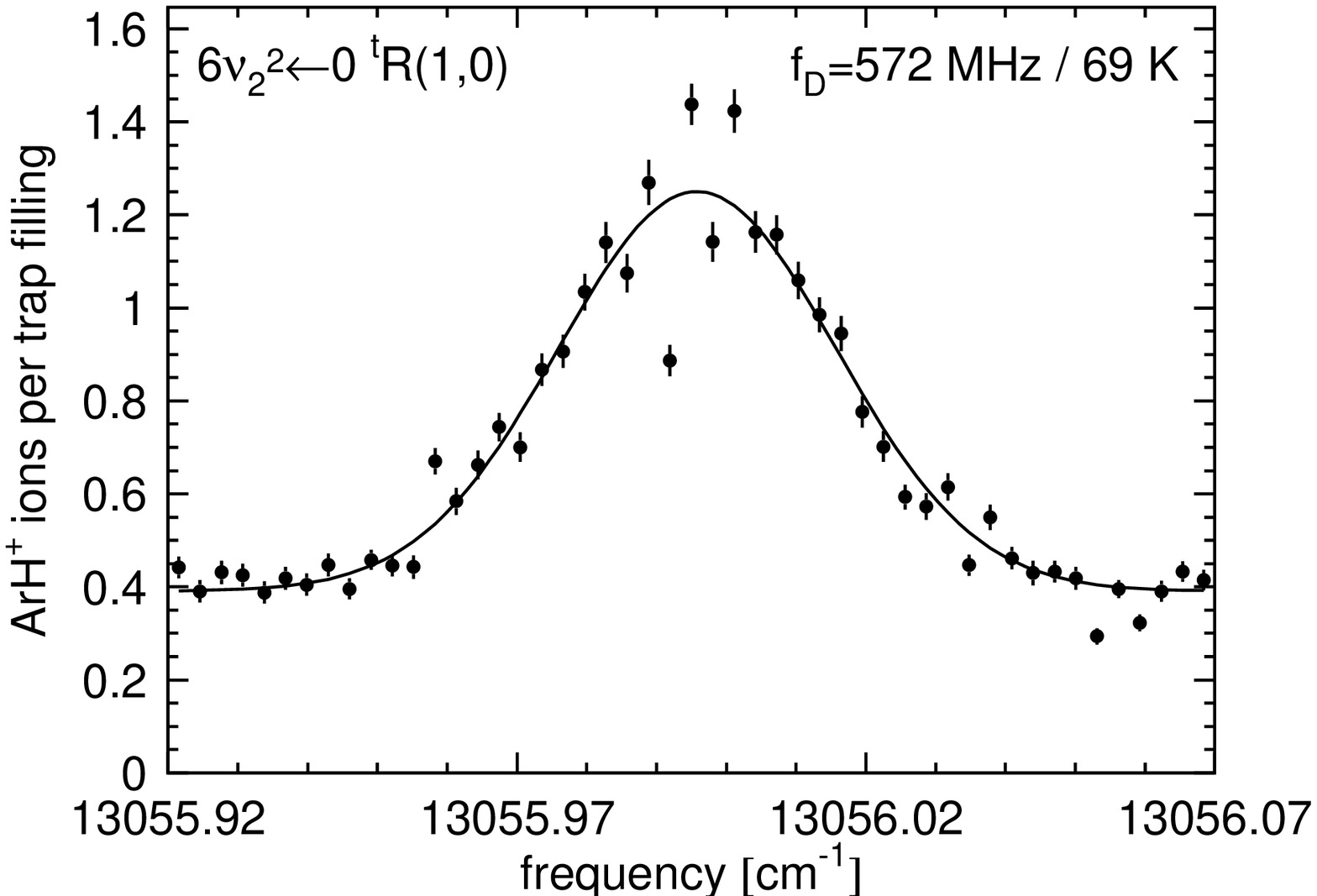}
\end{minipage}
\vspace*{-0.3cm}

\caption{Four exemplary spectra obtained by chemical probing spectroscopy. The translational temperature derived from the Dopppler width $f_D$ is given in the insets.}
\label{pic:spectra}
\end{figure}

After unloading the trap the ArH$^+$ ions are mass-selected by a quadrupole and counted by the Daly detector. The exact measurement procedure was adapted to the signal strength and scan width for each individual line. Routinely, 20 to 100 trap fillings were taken at a fixed laser wavelength. Afterwards the laser power and wavelength were determined and the wavelength was increased for the next step until a full scan cycle was completed. After each scan the average number of stored H$_3^+$ ions and the ArH$^+$ lifetime was measured. Typically 5 to 20 scans were acquired for each transition.  

\section{Results}

We have measured 23 transitions to H$_3^+$ states above the barrier to linearity, six of which had been previously observed by the Oka group and published by Gottfried \cite{gottfried06}. Transitions to the following vibrational bands have been observed: $5\nu_2$, $3\nu_1+1\nu_2$, $1\nu_1+4\nu_2$, $6\nu_2$,$2\nu_1+3\nu_2$ and $3\nu_1+2\nu_2$. For the transition assignment, information from several sources had to be combined. The NMT calculation provides the energy of the upper and lower level as well as their respective $J$ quantum numbers, but not the complete rovibrational assigment. With the lower level energy given in the NMT tables, its assignment is trivial since essentially only the two lowest H$_3^+$ levels are populated inside the trap. For the assignment of the upper level it has to be identified in the SAH publication \cite{schiffels03b} where all vibrational quantum numbers as well as $J$ and $G$ are specified. However, from the present work only those transitions to vibrational states with $5\nu_2$, $3\nu_1+1\nu_2$ and $1\nu_1+4\nu_2$ can be found in the original SAH publication, while for the other states with $6\nu_2$, $2\nu_1+3\nu_2$ and $3\nu_1+2\nu_2$ new level assignments were provided by A. Alijah \cite{alijah:private}. After all quantum numbers of the individual levels were found from these references, the transitions were labeled according to the recommendations given in \cite{lindsay01}. In Table\,\ref{table1} the rovibrational assignment for each transition is given together with the results of Gottfried (where available) and the spectroscopically adjusted transition energies according to SAH \cite{alijah:private}.

The experimental uncertainty is 0.01\,cm$^{-1}$ for all transitions, as determined by the absolute accuracy of the frequency measurement. The transition energies obtained by the Oka group agree with our new measurements within the combined uncertainties, however, they tend to be somewhat higher.   

In Table\,\ref{table2} the experimental transition energies are compared to the calculations of SAH and NMT. The findings of previous publications are confirmed, namely that the unadjusted calculation of SAH systematically overestimates the transition energies by up to 1\,cm$^{-1}$. In the first publication of SAH \cite{schiffels03a}, a method is developed that accounts for the effects of non-adiabatic coupling in order to adjust the energies of highly excited states to spectroscopic observations below 9\,000\,cm$^{-1}$. Here, we list comparisons with values including the $E^4$ correction (see \cite{schiffels03a} for details) and this correction rectifies some of the discrepancy -- especially at energies below $\sim$11800\,cm$^{-1}$ -- but it overshoots and at higher energies the discrepancy is as large as for the unadjusted values, albeit with opposite sign. Tuning the correction parameters to higher lying states might bring further improvement, once enough experimental data are available {\cite{alijah:private}.

 The line list calculation of NMT also shows a tendency to overestimate the transition energy with the scatter being generally larger than that of SAH. In the last column the calculated Einstein B$_{12}$ coefficients (based upon the Einstein $A_{21}$ coefficients of NMT) are shown as a measure of the transition strength. Comparison of the theoretical B$_{12}$ values for these lines to those of the bending mode fundamental $1\nu_2^1\leftarrow 0$, Q(1,0) listed in the last line of Table\,\ref{table2} reveals that some of the measured transitions are expected to be weaker by more than 5 orders of magnitude, which suggests them to be - to the best of our knowledge - the weakest H$_3^+$ transitions observed to date.

From the Doppler width of the distribution, the translational temperature of the ions can be derived. The natural linewidth as well as any broadening effects are negligible, hence the measured spectra feature perfect Gaussian line shapes (see Fig.\,\ref{pic:spectra}). A translational temperature around $\sim$70\,K is found for all of the lines, somewhat higher than the temperature of the trap walls. The reason for the discrepancy are probably imperfections in the 22-pole geometry leading to radiofrequency heating in the outer parts of the trap, close to the rods. In a previous study \cite{mikosch04} a much higher value ($\sim$130\,K) was found at the same nominal trap temperature. This large deviation was found to be dependent on the helium buffer gas density inside the trap which had been limited in the previous experiment due to impurities in the gas supply lines. In the present work stabilization of the translational temperature was found at higher helium densities and further increase did not result in lower translational temperatures.

\clearpage

\begin{table}[t]
\centering
\begin{tabular}{rcccc}
\hline
 & &\hspace{0.8cm} this work\hspace{1cm}  & \hspace{0.3cm}  Gottfried\,\cite{gottfried06} \hspace{0.3cm} &\hspace{0.7cm}  SAH$_{\rm ad}$\,\cite{alijah:private} \hspace{0.7cm} \\
\hspace{0.5cm} \raisebox{2.1ex}{vibration}\hspace{0.5cm}&\hspace{0.5cm} \raisebox{2.1ex}{rotation}\hspace{0.5cm} &   \raisebox{1.2ex}{exp. [cm$^{-1}$]} &  \raisebox{1.2ex}{exp. [cm$^{-1}$]} & \raisebox{1.2ex}{theory  [cm$^{-1}$]} \\
\hline\hline
$5\nu_2^1\leftarrow 0$ & $R(1,0)$   &  11\,228.598(10) & 11\,228.613(10) & 11\,228.54\\[-0.15cm]
$5\nu_2^1\leftarrow 0$ & \hspace*{0.1cm} $R(1,1)^u$ &  11\,244.350(10) & 11\,244.366(10) & 11\,244.34\\[0.05cm]
$3\nu_1+1\nu_2^1\leftarrow 0$ & $P(1,1)$ &  11\,258.975(10) &  & 11\,258.48\\[-0.15cm]
$3\nu_1+1\nu_2^1\leftarrow 0$ & $Q(1,1)$ &  11\,342.587(10) &  & 11\,342.11\\[-0.15cm]
$3\nu_1+1\nu_2^1\leftarrow 0$ & \hspace*{0.1cm} $R(1,1)^l$ &  11\,465.505(10) &  & 11\,464.95\\[-0.15cm]
$3\nu_1+1\nu_2^1\leftarrow 0$ & \hspace*{0.1cm} $R(1,1)^u$ &  11\,511.373(10) &  & 11\,510.89\\[0.05cm]
$5\nu_2^5\leftarrow 0$ & $^{-6}P(1,1)$ \hspace*{0.2cm} &  11\,594.276(10) &  & 11\,593.87\\[-0.15cm]
$5\nu_2^5\leftarrow 0$ & $^{-6}R(1,1)$ \hspace*{0.2cm} &  11\,707.257(10) & 11\,707.265(10)\footnote{For the $5\nu_2^5\leftarrow 0$ line found at 11 707.257\,cm$^{-1}$
the assigned upper quantum number is $G=5$, which for the lower
$G=1$ calls for a line type prefix of $-6$ unlike the one given in Ref.
\cite{gottfried06}.}& 11\,707.11\\[-0.15cm]
$5\nu_2^5\leftarrow 0$ & $^{-6}Q(1,1)$ \hspace*{0.2cm} &  11\,707.797(10) &  & 11\,707.38\\[0.05cm]
$1\nu_1+4\nu_2^0\leftarrow 0$ & $^nR(1,1)$ \hspace*{0.05cm} &  11\,882.376(10) &  & 11\,881.76\\[0.05cm]
$1\nu_1+4\nu_2^2\leftarrow 0$ & $^tQ(1,0)$ \hspace*{0.05cm} &  12\,018.812(10) &  & 12\,018.18\\[-0.15cm]
$1\nu_1+4\nu_2^2\leftarrow 0$ & $^tR(1,1)$ \hspace*{0.05cm} &  12\,086.738(10) &  & 12\,086.36\\[0.05cm]
$2\nu_1+3\nu_2^1\leftarrow 0$ & $P(1,1)$ &  12\,239.242(10) &  & 12\,238.65\\[-0.15cm]
$2\nu_1+3\nu_2^1\leftarrow 0$ & $Q(1,1)$ &  12\,373.325(10) &  & 12\,373.01\\[-0.15cm]
$2\nu_1+3\nu_2^1\leftarrow 0$ & $R(1,1)$ &  12\,381.135(10) &  & 12\,380.04\\[0.05cm]
$6\nu_2^2\leftarrow 0$ & $^nP(1,1)$ \hspace*{0.05cm} &  12\,413.257(10) &  & 12\,412.73\\[-0.15cm]
$6\nu_2^2\leftarrow 0$ & $^tQ(1,0)$ \hspace*{0.05cm} &  12\,419.127(10) & 12\,419.121(10)  & 12418.41\\[-0.15cm]
$6\nu_2^2\leftarrow 0$ & $^nQ(1,1)$ \hspace*{0.05cm} &  12\,623.160(10) &  & 12\,622.60\\[-0.15cm]
$6\nu_2^2\leftarrow 0$ & $^tR(1,1)$ \hspace*{0.05cm} &  12\,678.683(10) &  & 12\,678.08\\[0.05cm]
$1\nu_1+4\nu_2^4\leftarrow 0$ & $^tR(1,0)$ \hspace*{0.05cm} &  12\,897.877(10) & 12\,897.888(10)  & 12\,896.99\\[-0.15cm]
$6\nu_2^2\leftarrow 0$ & $^tR(1,0)$ \hspace*{0.05cm} &  13\,055.994(10) & 13\,056.013(10) & 13\,055.17\\[-0.15cm]
$2\nu_1+3\nu_2^3\leftarrow 0$ & $ R(1,1)$ \hspace*{0.05cm} &  13\,071.590(10) &  & 13\,070.84\\[-0.15cm]
$3\nu_1+2\nu_2^2\leftarrow 0$ & $^tR(1,1)$ \hspace*{0.05cm} &  13\,332.884(10) &  & 13\,331.70\\
\hline\hline
\end{tabular}
\caption{List of observed transitions together with their vibrational and rotational assignments using Refs.\,\cite{alijah:private, schiffels03b}. The transition frequencies are given for the present work as well as those published by Gottfried \cite{gottfried06} (where available) and the spectroscopically adjusted values of SAH \cite{alijah:private}.}
\label{table1}  
\end{table}

\begin{table}[h]
\begin{tabular}{cccccc}
\hline
 \hspace{0.4cm} observed\hspace{0.4cm} &\hspace{0.15cm}  SAH \cite{schiffels03b} \hspace{0.15cm} & \hspace{0.5cm}  SAH$_{\rm ad}$ \cite{alijah:private} \hspace{0.15cm} & \hspace{0.15cm} NMT\cite{neale96}\hspace{0.15cm} &\hspace{0.65cm} B$_{12}$\cite{neale96}\hspace{0.65cm} \\[-0.17cm]
[cm$^{-1}$] &  [cm$^{-1}$] &  [cm$^{-1}$]  & [cm$^{-1}$] & [cm$^3$/Js$^{2}$]\\
\hline\hline
11\,228.598(10) & 1.08 & -0.06 & 0.25 & 7.20e+19 \\[-0.15cm] 
11\,244.350(10) & 1.13 & -0.01 & 0.23 & 3.32e+19 \\[0.05cm]
11\,258.975(10) & 0.66 & -0.50 & 1.59 & 4.96e+18 \\[-0.15cm] 
11\,342.587(10) & 0.67 & -0.48 & 1.61 & 6.79e+18 \\[-0.15cm] 
11\,465.505(10) & 0.62 & -0.56 & 1.67 & 6.11e+18 \\[-0.15cm] 
11\,511.373(10) & 0.69 & -0.48 & 1.62 & 3.54e+18 \\[0.05cm]
11\,594.276(10) & 0.77 & -0.41 & 0.88 & 6.76e+18 \\[-0.15cm] 
11\,707.257(10) & 1.04 & -0.15 & 1.02 & 1.62e+19 \\[-0.15cm] 
11\,707.797(10) & 0.77 & -0.42 & 1.06 & 1.16e+19 \\[0.05cm]
11\,882.376(10) & 0.59 & -0.62 & 0.95 & 4.65e+18 \\[0.05cm]
12\,018.812(10) & 0.60 & -0.63 & 0.95 & 2.46e+18 \\[-0.15cm] 
12\,086.738(10) & 0.85 & -0.38 & 0.71 & 7.71e+18 \\[0.05cm]
12\,239.242(10) & 0.65 & -0.59 & 0.23 & 6.83e+18 \\[-0.15cm] 
12\,373.325(10) & 0.95 & -0.32 & 1.20 & 4.06e+18 \\[-0.15cm] 
12\,381.135(10) & 0.17 & -1.10 & 0.57 & 4.11e+18 \\[0.05cm] 
12\,413.257(10) & 0.73 & -0.53 & -0.27 & 3.74e+18 \\[-0.15cm] 
12\,419.127(10) & 0.55 & -0.72 & 0.19 & 1.21e+19 \\[-0.15cm] 
12\,623.160(10) & 0.72 & -0.56 & 0.80 & 9.05e+18 \\[-0.15cm] 
12\,678.683(10) & 0.69 & -0.60 & 0.84 & 8.00e+18 \\[0.05cm]
12\,897.877(10) & 0.42 & -0.89 & 2.20 & 7.46e+18 \\[-0.15cm] 
13\,055.994(10) & 0.51 & -0.82 & 0.64 & 1.20e+19 \\[-0.15cm] 
13\,071.590(10) & 0.58 & -0.75 & 0.41 & 6.99e+18 \\[-0.15cm] 
13\,332.884(10) & 0.18 & -1.18 & 2.85 & 2.03e+18 \\
\hline
$1\nu_2^1$ fundamental \\[-0.15cm]
\hspace*{0.5cm}2529.724(05) \cite{mckellar98} & 0.23 & -0.04 & 0.01 & 4.77e+23 \\
\hline\hline 
\end{tabular}
\caption{List of the observed transition frequencies and deviations to the unadjusted SAH calculation and corrected values of SAH$_{\rm ad}$ \cite{alijah:private} and the NMT calculation. The Einstein B$_{12}$ coefficients derived from NMT \cite{neale96} are also given. In the last line the parameters of the fundamental bending mode transition $1\nu_2^1\leftarrow 0$, Q(1,0) are shown for comparison.}
\label{table2}      
\end{table}

\clearpage

\section{Perspectives}

The results presented in this work demonstrate that chemical probing spectroscopy can be used to perform measurements of weak rovibrational transitions with unprecedented efficiency as well as very low background. The sensitivity of this approach is currently limited by the number of H$_3^+$ ions inside the trap and by the lifetime of ArH$^+$. While for the present study only a few thousand H$_3^+$ ions were stored, up to 2$\times$10$^6$ ions can be stored inside the trap when it is used as an ion injector at 10\,K \cite{kreckel05}. Although, at the more demanding conditions necessary for spectroscopic measurements, this number will most likely not be achieved entirely, the practical sensitivity limit for this type of investigations clearly has not yet been reached in the present work and is expected to be revealed only by further studies.

\begin{table}[b]
\centering
\begin{tabular}{cccccc}
\hline
 \hspace{0.1cm} frequency \hspace{0.1cm} &  \hspace{0.1cm} wavelength \hspace{0.1cm}  & \hspace{0.5cm}  A$_{21}$ (MBB)  \hspace{0.5cm}   &  \hspace{0.4cm}   B$_{12}$ (MBB)  \hspace{0.4cm}  & \hspace{0.5cm}  A$_{21}$ (JS)  \hspace{0.5cm}   &  \hspace{0.4cm}   B$_{12}$ (JS)  \hspace{0.4cm}  \\[-0.25cm]
  [cm$^{-1}$] &  [nm]  &  [s$^{-1}$] & [cm$^3$/Js$^{2}$]  &  [s$^{-1}$] & [cm$^3$/Js$^{2}$] \\ 
\hline\hline
2\,521 & 3\,967 & 1.29e+02 & 4.83e+23 & 1.20e+02 & 4.52e+23 \\[-0.25cm] 
4\,997 & 2\,001 & 1.49e+02 & 7.19e+22 & 1.68e+02 & 8.06e+22 \\[-0.25cm] 
7\,003 & 1\,428 & 1.65e+01 & 2.88e+21 & 3.79e+01 & 6.63e+21 \\[-0.25cm] 
9\,108 & 1\,098 & 6.55e+00 & 5.21e+20 & 1.26e+01 & 1.00e+21 \\[-0.25cm] 
10\,853 & 921 & 2.17e+00 & 1.02e+20 & 5.20e+00 & 2.44e+20 \\[-0.25cm] 
12\,294 & 813 & 6.70e-01 & 2.17e+19 & 1.40e+00 & 4.52e+19 \\[-0.25cm] 
13\,681 & 731 & 5.60e-01 & 1.31e+19 & 1.60e+00 & 3.75e+19 \\[-0.25cm] 
15\,103 & 662 & 2.90e-01 & 5.05e+18 & 9.60e-01 & 1.67e+19 \\[-0.25cm] 
16\,712 & 598 & 1.30e-01 & 1.67e+18 & 2.30e-01 & 2.96e+18 \\[-0.25cm] 
18\,432 & 543 & 6.00e-02 & 5.75e+17 & 1.40e-01 & 1.34e+18 \\[-0.25cm] 
20\,238 & 494 & 2.50e-02 & 1.81e+17 & 6.40e-02 & 4.64e+17 \\[-0.25cm] 
22\,115 & 452 & 1.10e-02 & 6.11e+16 & 2.00e-02 & 1.11e+17 \\[-0.25cm] 
24\,034 & 416 & 7.70e-03 & 3.33e+16 & 1.40e-02 & 6.06e+16 \\[-0.25cm] 
25\,912 & 386 & 3.30e-03 & 1.14e+16 & 5.60e-03 & 1.93e+16 \\[-0.25cm] 
27\,803 & 360 & 1.80e-03 & 5.03e+15 & 3.70e-03 & 1.03e+16 \\ 
\hline\hline 
\end{tabular}
\caption{List of the brightest vibrational bands of H$_3^+$ up to 28\,000\,cm$^{-1}$, from a calculation of Le Sueur {\it et al.} \cite{lesueur93}. The Einstein $A_{21}$ coefficients are those to the ground state. Two different potential energy surfaces have been employed, MBB \cite{meyer86} and JS \cite{jensen86}, respectively.} 
\label{table3}      
\end{table}

Apart from the benefit from higher H$_3^+$ numbers as seen in eq.\,(\ref{eq:formation}), eq. (\ref{eq:diffsolve}) shows that the number of ArH$^+$ ions produced for long laser interaction times is also directly proportional to the ArH$^+$ lifetime. In order to improve this lifetime -- limited by the residual H$_2$ number density in the trapping region as discussed in Sec.\,IIIB -- an upgrade introducing a differential pumping stage between the ion source and the cryogenic trap is in progress.

It is interesting to consider the potential of the chemical probing technique in view of the calculated H$_3^+$ transition strengths for the visible- or even the UV-range. Le Sueur {\it et al.} \cite{lesueur93} have calculated the strengths of the most effective H$_3^+$ bending mode transitions (accessing the so-called horseshoe states) up to the dissociation limit. In Table\,\ref{table3} the Einstein $A_{21}$ coefficients for the vibrational bands up to 28\,000\,cm$^{-1}$ from this calculation are shown and converted to Einstein $B_{12}$ coefficients which are better suited to estimate the feasibility of spectroscopic studies. Two different potential energy surfaces were used for the calculations. Both predict that the $B_{12}$ coefficients drop off by roughly an order of magnitude at $\sim$16\,700\,cm$^{-1}$ and two orders of magnitude at  $\sim$20\,200\,cm$^{-1}$, when compared to the present measurements around $\sim$12\,300\,cm$^{-1}$. All of the transitions in Table\,\ref{table3} would be in the range of tunable dye lasers, which can provide similar output power as the Ti:Sa laser used in the present experiment. Hence, with the high potential of the chemical probing technique, their observation appears possible, although the demand on the accuracy of the theoretical predictions will increase in these new regimes to restrict the required scan range for the measurements. Nevertheless, with the ongoing upgrade of the 22-pole setup, an extension of H$_3^+$ rovibrational spectroscopy into the visible or dye laser regime seems feasible in the near future.     

\section*{Acknowledgement}
We thank A. Alijah for providing new calculations and revised assignments prior to publication. We are also indebted to J. Tennyson for fruitful discussions. We thank M. Kowalska, R. Neugart and G. Huber for providing the Ti:Sa laser for the experiment. H.K. was supported in part by NSF Astronomy and Astrophysics Research Grant AST-0606960.

\bibliography{kreckel_h3tisa}

\end{document}